\documentclass[aip, amsmath, amssymb, graphicx, multirow]{revtex4-1}

\usepackage{graphicx}
\usepackage{amsmath}
\usepackage{amsfonts}
\usepackage{tabularx}
    \newcolumntype{L}{>{\raggedright\arraybackslash}X}
\usepackage{dcolumn}
\usepackage{float}
\usepackage{multirow}

\draft % marks overfull lines with a black rule on the right

\begin{document}

\title{On the complex solution of the Schr\"odinger equation with exponential potentials}

\author{Javier Garcia}
\email[]{jgarcia@fisica.unlp.edu.ar}
\affiliation{$^1$Instituto de F\'isica La Plata, Consejo Nacional de Investigaciones Cient\'ificas y T\'ecnicas, and Departamento de F\'isica, Universidad Nacional de La Plata, C.C. 67, 1900 La Plata, Argentina}

\date{\today}

\begin{abstract}
    We study the analytical solutions of the Schr\"odinger equation with a repulsive exponential potential $\lambda e^{- r}$, and that with an exponential wall $\lambda e^r$, both with $\lambda > 0$. 
    We show that the complex eigenenergies obtained for the latter tend either to those of the former, or to real rational numbers as $\lambda \rightarrow \infty$.
    In the light of these results, we explain the wrong resonance energies obtained in a previous application of the Riccati-Pad\'e method to the Schr\"odinger equation with a repulsive exponential potential, and further study the convergence properties of this approach.
\end{abstract}


\maketitle %\maketitle must follow title, authors, abstract and \pacs

\section{Introduction}

Resonances are solutions of the Schr\"odinger equation that have complex energies and present a purely outgoing behavior in the asymptotic region. These states, introduced by Siegert to describe decaying nuclear states using a stationary-state picture\cite{Siegert_1939}, arise, together with bound and virtual states (i.e., states that grow exponentially in the asymptotic region), as poles of the scattering matrix.
They spark a lot of interest among physicists and chemists, evidenced by the large amount of methods devised for their computation.
Most of them are based on the complex rotation method, introduced by Aguilar, Balslev, and Combes \cite{Aguilar_1971, Balslev_1971}, which consists in performing a dilation on the coordinate $U(\theta) r U^{-1}(\theta) = e^{i\theta}r$, by choosing the parameter $\theta$ in such a way that the outgoing-wave state is transformed into one that asymptotically decays to zero, as bound states do, and then using one of the many numerical methods devised to compute the latter.
The first applications of the complex rotation method involved a variational approach, which remains popular (e.g. Refs. \onlinecite{Fernandez_2013_a, Kuro__2015, Myo_2020}) but also different techniques have been employed, such as numerical integration methods \cite{Rittby_1981, Rittby_1982, Atabek_1982, Atabek_1983}, dimensional scaling \cite{Kais_1993}, the Lanczos' Tau method \cite{Midy_1993}, spherical-box approaches \cite{Maier_1980, Zhou_2009}, direct computation of the Jost function \cite{Sofianos_1997}, gradient optimization \cite{Bai_2021}, and eigenvector continuation \cite{Yapa_2023}, among others.
There is also a wide array of methods that do not involve a complex rotation; examples of these include methods based on Siegert pseudo-states \cite{Tolstikhin_1997, Tolstikhin_1998, Batishchev_2007,curik_2023}, the complex absorbing potential \cite{Riss_1993, M_ller_2003}, Pad\'e approximations of the S-matrix \cite{Rakityansky_2007}, the real stabilization method \cite{Zhang_2008}, the coupled channels method \cite{Liang_2015}, and the Riccati-Pad\'e method (RPM) \cite{Fernandez_1995, Fern_ndez_1995}.
The list of methods provided here is by no means comprehensive.

The RPM consists in expanding a modified logarithmic derivative of the wavefunction in a Taylor series, and constructing Hankel determinants with the expansion coefficients, the roots of which give rise to sequences that converge rapidly towards both the bound states and resonances \cite{Fernandez_1989, Fernandez_1989_a}.
It was originally proposed for the computation of bound states, but later it was found that it is also able to yield resonances without resorting explicitly to a complex rotation \cite{Fernandez_1995, Fern_ndez_1995}.
Several applications to the computation of resonances ensued \cite{Fern_ndez_1996, Fern_ndez_1997, Fern_ndez_2008, Amore_2008, Fern_ndez_2012, Fernandez_2016_a, Fernandez_2018}, which showed that it is able to compute them very accurately and with comparatively little cost. 
For example, the resonances for the Stark effect in the hydrogen atom computed by us in Ref. \onlinecite{Fernandez_2018} are to our knowledge the most accurate available in literature.

As with bound states, there are very few systems whose resonances can be computed analytically.
One of such systems are the $s-$states of a particle exposed to a decaying exponential potential, i.e., $V(r) = \lambda e^{-r}$, with real $\lambda$.
If $\lambda < 0$, the system admits a finite number of bound states \cite{Ma_1946, Ter_Haar_1946}. 
On the other hand if $\lambda > 0$, the potential is not expected to hold bound or resonant states, but some of the eigenfunctions, which correspond to complex eigenvalues, behave as such \cite{Atabek_1982}.
In addition, the repulsive exponential potential has a set of virtual states.
The authors of Ref. \onlinecite{Atabek_1982} studied the complex eigenstates of the complex exponential potential both analytically and by means of the complex rotation method, and soon after the eigenvalues were used as benchmarks to test a few of the other numerical methods mentioned in the present work \cite{Kais_1993, Midy_1993, Fernandez_1995, Fern_ndez_1996}.
Among these is the RPM, which produced some baffling results: it yields complex energies that are very close to those of the analytical solutions, but are not quite the same, unlike the approaches of Refs. \onlinecite{Atabek_1982,Kais_1993,Midy_1993}, which yielded the correct ones.
The wrong results provided by the RPM were later confirmed in a more systematic study of the same problem \cite{Amore_2008}.
Ref. \onlinecite{Fern_ndez_1996} also saw wrong results for the Schr\"odinger equation with the one-dimensional potential $V(x) = (x^2-2J) \exp(-\lambda x^2)$, introduced by Moiseiev to model pre-dissociation resonances of diatomic molecules \cite{Moiseyev_1978},
but it was later explained in Ref. \onlinecite{Fernandez_2016_a} that the incorrect eigenvalues can also be obtained by complex rotation, provided that the rotation angle is set to be greater than the critical value $\theta_{{\rm crit}} = \pi/4$. 
In fact, such eigenvalues had already been obtained and discussed by other authors \cite{Rittby_1981, Rittby_1982, Korsch_1982, Rittby_1982_a}.
It is reasonable to expect that the discrepancy between the exact complex eigenenergies for the repulsive exponential potential and those obtained by the RPM can be explained in a similar fashion to those for the potential introduced in Ref. \onlinecite{Moiseyev_1978}.
In the present work, we show that this assumption is correct, and that the seemingly wrong resonances obtained by means of the RPM are in fact complex solutions of the Schr\"odinger equation for a particle exposed to an infinite exponential well, $V(r) = \lambda e^r$.
Even though the latter has been solved analytically\cite{Atabek_1983}, to our knowledge, both spectra have not yet been compared.

This work is organized as follows.
In Sec. \ref{sec:analytical}, we review the analytical solution of the Schr\"odinger equation for both problems, and we perform a comparison between their eigenspectra.
Then, in Sec. \ref{sec:RPM}, we employ an efficient implementation of the RPM to improve upon the computations of Refs. \onlinecite{Fernandez_1995,Fern_ndez_1996,Amore_2008}, and we study the rate of convergence of the roots of the Hankel determinants towards both kinds of complex eigenvalues.
We finally sum up our results and draw further conclusions in Sec. \ref{sec:conclusions}.

\section{Analytical treatment} \label{sec:analytical}

\subsection*{The Schr\"odinger equation with a repulsive exponential potential}

We first treat the Schr\"odinger equation with a repulsive exponential potential, 
\begin{equation}
    -\phi^{\prime\prime}(r) + \lambda e^{-r} \phi(r) = \varepsilon \phi(r), 
    \label{eq:schrodinger}
\end{equation}
with $\lambda > 0$ and $\phi(0) = 0$.
As discussed in the Introduction, this problem was already studied in Ref. \onlinecite{Atabek_1982} in great detail, but for consistency, we briefly describe its analytical solution in the following paragraphs.

By defining $\mu = \sqrt{-4\varepsilon}$, Eq. \eqref{eq:schrodinger} is exactly solvable in terms of the Bessel functions of the first kind $J_{\pm \mu}(z)$, where $z = 2\sqrt{-\lambda} e^{-r/2}\,$ \cite{Ma_1946, Atabek_1982}, or, equivalently, in terms of the modified Bessel functions of the first kind $I_{\pm\mu}(t)$, where $t = 2 \sqrt{\lambda} e^{-r/2}$.
We prefer the latter since for real $r$, $t$ is also real, whereas $z$ is imaginary.
The general solution of Eq. \eqref{eq:schrodinger} can be written as $A_\mu I_\mu(t) + B_\mu I_{-\mu}(t)$, and the condition that $\phi(r)$ behaves as an outgoing wave, i.e., $\phi(r) \sim e^{ikr}$, with $k = i\mu/2$, ${\rm Re}(k) > 0$ and ${\rm Im}(k) < 0\,$ \cite{Atabek_1982}, which implies ${\rm Re}(\mu) < 0$ and ${\rm Im}(\mu) < 0$, requires that $B_\mu = 0$, since for small $|t|$, $I_{\pm \mu}(t) \sim (t/2)^{\pm \mu} / \Gamma(\mu+1)$.
The condition $\phi(0) = 0$ results in
\begin{equation}
    I_{\mu_n}(2\sqrt{\lambda}) = 0,
    \label{eq:true_resonance}
\end{equation}
which yields the eigenvalues $\varepsilon_n = -\mu_n^2/4$.
Here $n = 0, 1, \ldots$ serves as a label that orders them by increasing absolute value.
It is well known that the solutions of Eq. \eqref{eq:true_resonance} are either real and negative or come in complex conjugate pairs, with ${\rm Re}(\mu) < 0$ \cite{Atabek_1982}.
For $\lambda \rightarrow 0$, they tend to negative integers, and as $\lambda$ increases, pairs of solutions coalesce to form a complex pair, one pair at a time.
The solutions with real $\mu < 0$ correspond to virtual states, i.e., states that grow exponentially as $r \rightarrow \infty$, and the complex solutions correspond to either resonances, if ${\rm Im}(\mu) < 0$, or growing states ${\rm Im}(\mu) > 0$.
Since the roots of Eq. \eqref{eq:true_resonance} come in complex conjugate pairs, every resonance comes with an associated growing state.
In the present work, the notation $\mu_n$ will refer only to resonances. 
If needed, the associated growing states will be referred to by using complex conjugation, i.e. $\mu_n^*$.

The authors of Ref. \onlinecite{Atabek_1982} were limited to solving Eq. \eqref{eq:true_resonance} for relatively small values of $\lambda$, since it was easier for them to find its roots by means of the computational resources that were available at that time.
Contrarily, nowadays they are quite trivial to find with any number of significant digits by using any modern computer algebra software, or a multiprecision library such as mpmath \cite{mpmath}, which we have used extensively in the present work.
%The authors of Ref. \cite{Atabek_1982} also utilized the complex rotation method combined with two different numerical integration methods to obtain the eigenvalues of Eq. \eqref{eq:schrodinger} for a few small values of $\lambda$, showing that they coincide with the solutions of Eq. \eqref{eq:true_resonance}.

\subsection*{The Schr\"odinger equation with an infinite exponential well}

We now focus our attention on the Schr\"odinger equation with an infinite exponential well potential, i.e., 
\begin{equation}
    -\phi^{\prime\prime}(r) + \lambda e^r \phi(r) = \epsilon \phi(r).
    \label{eq:schrodinger2}
\end{equation}
It is related to Eq. \eqref{eq:schrodinger} by a change of variable $r \rightarrow -r$, and it is also solvable in terms of Bessel functions, as was shown by Atabek and Lefebvre in Ref. \onlinecite{Atabek_1983}. 
They posed Eq. \eqref{eq:schrodinger2} as a three-parameter problem, with $V(r) = A e^{\alpha(r - r_0)}$, but by making the change of variable $r \rightarrow r /\alpha$, setting $\lambda = A \exp\left(-\alpha r_0\right)/\alpha^2$, and multiplying the energy by $\alpha^2$, Eq. \eqref{eq:schrodinger2} is obtained without loss of generality.
In the present work we would like to provide an equivalent derivation of its analytical solution that we deem more adequate for the ensuing discussion.

The general solution of Eq. \eqref{eq:schrodinger2} can be written $A_\nu I_\nu(s) + B_\nu K_\nu(s)$, where $s = 2\sqrt{\lambda}e^{r/2}$, and $K_\nu$ is the modified Bessel function of the second kind, defined for noninteger $\nu$ as $K_\nu(s) = \pi \csc(\nu\pi) [I_{-\nu}(s) - I_{\nu}(s)]/2$, and for integer $n$, $K_n(s) = \lim_{\nu\rightarrow n} K_\nu(s)$.
It could also be written in terms of $I_{\pm \nu}(s)$, but we choose $K_\nu$ instead since for $|s| \rightarrow\infty$ they admit the following asymptotic expansions, 
\begin{align}
    I_\nu(s) &\sim \left( \frac{1}{2\pi s}\right)^{1/2} e^s \left[ 1 - \frac{4\nu^2-1}{8z} + \frac{(4\nu^2-1)(4\nu^2-9)}{2!(8z)^2} + \ldots \right] \label{eq:asymp_I}\\
    K_\nu(s) &\sim \left( \frac{\pi}{2s}\right)^{1/2} e^{-s}\left[ 1 + \frac{4\nu^2-1}{8z} + \frac{(4\nu^2-1)(4\nu^2-9)}{2!(8z)^2} + \ldots \right], \label{eq:asymp_K}.
\end{align}
Eq. \eqref{eq:asymp_I} is valid only when $-\pi/2 < {\rm arg}\,s < \pi$, whereas Eq. \eqref{eq:asymp_K} is valid in the range $-\pi < {\rm arg}\,s < \pi$.
For real $r$ (and, consequently, real $s$) the decaying eigenfunctions of Eq. \eqref{eq:schrodinger2} require $A_\nu = 0$, and the condition $\phi(0) = 0$ implies
\begin{equation}
    K_\nu(2\sqrt{\lambda}) = 0,
    \label{eq:bound_state}
\end{equation}
which yields the bound states.
For $\lambda>0$, there is an infinite amount of roots of Eq. \eqref{eq:bound_state}, and all of them are all imaginary \cite{Bateman_1953, Bagirova_2020, Krynytskyi_2021}, which implies that the spectrum is comprised of an infinite number of positive energies.
The eigenfunctions are strongly decaying for $r \rightarrow \infty$, since $\phi(r) \sim \sqrt{\pi/\sqrt{\lambda}} \exp[ -2 e^{r/2}\sqrt{\lambda}-r/4 ]$.

We now study the solutions to Eq. \eqref{eq:schrodinger2} under a complex rotation of the form $r = \rho e^{i\theta}$, with $(\rho, \theta) \in \mathbb{R}$.
We focus on the interval $(-\pi/2, \pi/2)$ since when $\theta = \pm \pi/2$, the asymptotic behavior of the solutions changes drastically, as $r$ becomes imaginary.
In fact, as $|\theta| > \pi/2$, the real part of $r$ changes sign, and the problem could be reinterpreted as that of Eq. \eqref{eq:schrodinger}, but with $r = \rho e^{i\theta^\prime}$, and $\theta^\prime = \pi - \theta$; we may therefore define $\theta_{\rm crit} = \pi/2$ as the critical complex rotation angle for which the infinite well is transformed into the finite repulsive exponential potential.
The complex-rotated $s$ can be written as
\begin{equation}
s = 2\sqrt{\lambda} e^{\frac{1}{2}\rho \cos \theta} e^{\frac{1}{2}i\rho \sin \theta},
\label{eq:s}
\end{equation}
and, since ${\rm arg}\,s = \rho \sin(\theta)/2$, it becomes apparent that not only does ${\rm arg}\,s$ change with $\theta$, but also with $\rho$.
When $|\rho \sin \theta| > \pi$, Eq. \eqref{eq:asymp_I} is no longer valid, and when $|\rho \sin \theta| > 2\pi$, Eq. \eqref{eq:asymp_K} becomes invalid, as well.
To circumvent this problem, we resort to the analytical continuation formulae for the modified Bessel functions \cite{dlmf},
\begin{align}
    I_v(s e^{m\pi i}) &= e^{m\nu\pi i}I_\nu(s) \label{eq:cont_I} \\
    K_v(s e^{m\pi i}) &= e^{-m\nu\pi i}K_\nu(s) - \pi i \sin(m\nu\pi) \csc(\nu\pi) I_\nu(s), \label{eq:cont_K}
\end{align}  
where $m = 0, \pm 1, \pm 2, \ldots$.
Eqs. \eqref{eq:cont_I} and \eqref{eq:cont_K} can be employed to extend the validity of Eqs. \eqref{eq:asymp_I} and \eqref{eq:asymp_K} to regions where ${\rm arg}\,s$ is greater than $\pi/2$.
For example, if $\rho$ and $\theta$ are such that $\pi/2 < {\rm arg}\,s < 3\pi/2$, then by setting $m = -1$ in Eqs. \eqref{eq:cont_I} and \eqref{eq:cont_K}, the argument of $s e^{m\pi i}$ stays within the region $(-\pi/2, \pi/2)$.
In general, for any given value of $\rho$, an integer $m$ can be chosen such that ${\rm arg}(s e^{m\pi i})$ lies between $-\pi/2$ and $\pi/2$, both for positive and negative $\theta$.
If $\lambda$ is large enough, the RHS of Eq. \eqref{eq:cont_K} is expected to behave as an asymptotically decaying function if
\begin{equation}
    -(1+2m)\pi < \rho \sin \theta < -(1-2m)\pi,
    \label{eq:convergence_region}
\end{equation}
where $m$ and $\theta$ have opposite signs.
Eq. \eqref{eq:bound_state} is now generalized,
\begin{equation}
    \begin{aligned}
    e^{-m{\nu_n^{(m)}}\pi i}K_{\nu_n^{(m)}}(2\sqrt{\lambda}) - \pi i \sin[m{\nu_n^{(m)}}\pi] \csc[{\nu_n^{(m)}}\pi] I_{\nu_n^{(m)}}(2\sqrt{\lambda}) = 0,
    \end{aligned}
    \label{eq:resonance}
\end{equation}
where again we use $n$ as a label for the eigenvalues, according to their increasing absolute value, and now the value of $m$ is added as a parenthesized superindex.
The energy of each state is computed as $\epsilon_n^{(m)} = -[\nu_n^{(m)}]^2/4$.
When $m = 0$ we recover Eq. \eqref{eq:bound_state}, but when $m \neq 0$, a different set of complex solutions is obtained.
Since for real $x$, $[I_\nu(x)]^* = I_{\nu^*}(x)$, and $[K_\nu(x)]^* = K_{\nu^*}(x)$, by taking the complex conjugate of Eq. \eqref{eq:resonance}, it can be readily seen that $\nu_n^{(-m)} = [\nu_n^{(m)}]^*$.

Eq. \eqref{eq:resonance} generalizes Eqs. (25), (28) and (32) of Ref. \onlinecite{Atabek_1983}.
The equivalence of both approaches is made more apparent by substituting the definition of $K_\nu$ in terms of $I_{\pm \nu}$ into Eq. \eqref{eq:resonance} and dividing by $e^{m\nu\pi i}$, whereas we obtain
\begin{equation}
    \csc (\pi\nu) \left [ e^{m\nu\pi i} I_{-\nu}(2\sqrt{\lambda}) - e^{-m\nu\pi i} I_\nu(2\sqrt{\lambda}) \right ] = 0,
    \label{eq:resonance2}
\end{equation}
[in the comparison, note that $I_{\nu}(z) = e^{\mp \nu \pi i/2} J_\nu(z e^{\pm \pi i/2})$].
The factor $\csc(\pi\nu)$ cannot be overlooked since it becomes singular when $\nu$ is an integer number.
It is also apparent from Eq. \eqref{eq:resonance2} that if $\nu_n$ is a solution, $-\nu_n$ is one as well, which is to be expected since only $\nu^2$ appears in the Schr\"odinger equation.
To simplify the upcoming discussions, we will refer to the resonances that correspond to the complex solutions of Eq. \eqref{eq:true_resonance} as ``barrier resonances'', (even though the repulsive exponential potential is not a barrier, the complex solutions of that problem behave as it were) and to those that correspond to Eqs. \eqref{eq:resonance} and \eqref{eq:resonance2} as ``well resonances''.
We may as well refer to Eq. \eqref{eq:schrodinger} as the ``barrier problem'', and to Eq. \eqref{eq:schrodinger2} as the ``well problem''.
The notation we employ to distinguish between the solutions of Eqs. \eqref{eq:schrodinger} and \eqref{eq:schrodinger2}, as well as the one we use to refer to the quantities computed in Sec. \ref{sec:RPM}, is summarized in Table \ref{tab:notation}.

\begin{table}[t]
    \begin{footnotesize}
        \begin{tabularx}{\linewidth}{lL}
            Symbol \hspace{1cm} & Meaning \\
            \hline
            $\mu_n$ & Solutions of Eq. \eqref{eq:true_resonance}, ordered by increasing absolute value. If $\mu_n$ is complex, ${\rm Im}\,\mu_n < 0$. \\
            $\nu_n^{(m)}$ & Solutions of Eq. \eqref{eq:resonance}, ordered by increasing absolute value. \\
            $\nu_{n^*}^{(m)}$ & Solution of Eq. \eqref{eq:resonance} that is closest to $\mu_n$. \\
            $\varepsilon_n$, $\epsilon_{n}^{(m)}$, $\epsilon_{n^*}^{(m)}$ & $-\mu_n^2/4$, $-[\nu_{n}^{(m)}]^2/4$, $-[\nu_{n^*}^{(m)}]^2/4$. \\
            $\tilde{\varepsilon}_n$, $\tilde{\epsilon}_{n}^{(m)}$, $\tilde{\epsilon}_{n^*}^{(m)}$ & Roots of $H_D^d$ closest to $\epsilon_n$, $\epsilon_{n}^{(m)}$, and $\epsilon_{n^*}^{(m)}$ . \\
            $\Delta_n, \Delta_n^{(m)}, \Delta_{n^*}^{(m)}$ & $-\log_{10} | \varepsilon_n - \tilde{\varepsilon}_n | $, $-\log_{10} | \epsilon_n^{(m)} - \tilde{\epsilon}_n^{(m)}|$, $-\log_{10} | \epsilon_{n^*}^{(m)} - \tilde{\epsilon}_{n^*}^{(m)}|$.
        \end{tabularx}
    \end{footnotesize}
    \caption{\label{tab:notation} Summary of notation.}
\end{table}

\subsection*{Comparison of both spectra}

\begin{figure}[t]
        \begin{center}
            \includegraphics[]{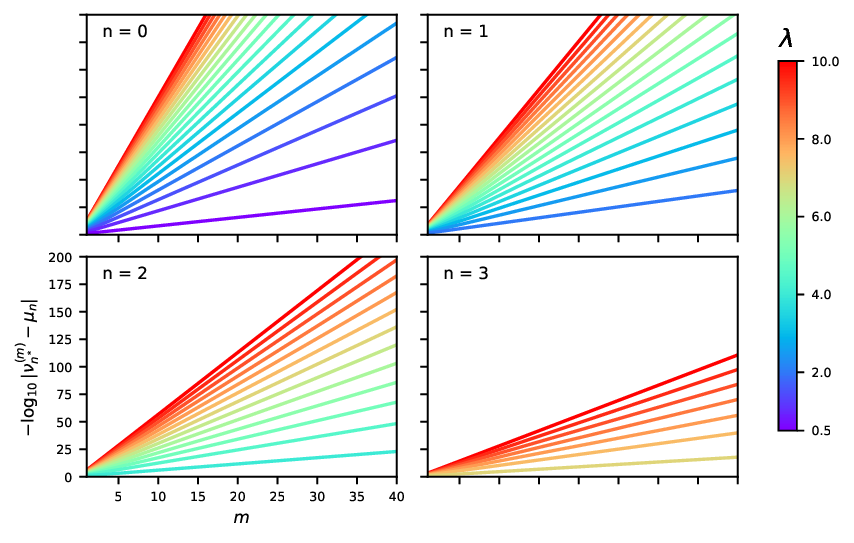}
            \caption{\label{fig:comparison} Logarithmic difference between the roots of Eqs.  \eqref{eq:true_resonance} and \eqref{eq:resonance}.}
        \end{center}
\end{figure}

We now focus on comparing the well and barrier resonances.
To that end, we have computed the well resonances for $\lambda$ ranging from $0.1$ to $12$, and $m = 1, 2, 3, 4$.
We have also computed the barrier resonances for the same values of $\lambda$; both kinds are shown in Fig. \ref{fig:comparison} (the barrier resonances are the same for each plot, since they do not depend on $m$).
\begin{figure}[t]
        \begin{center}
            \includegraphics[]{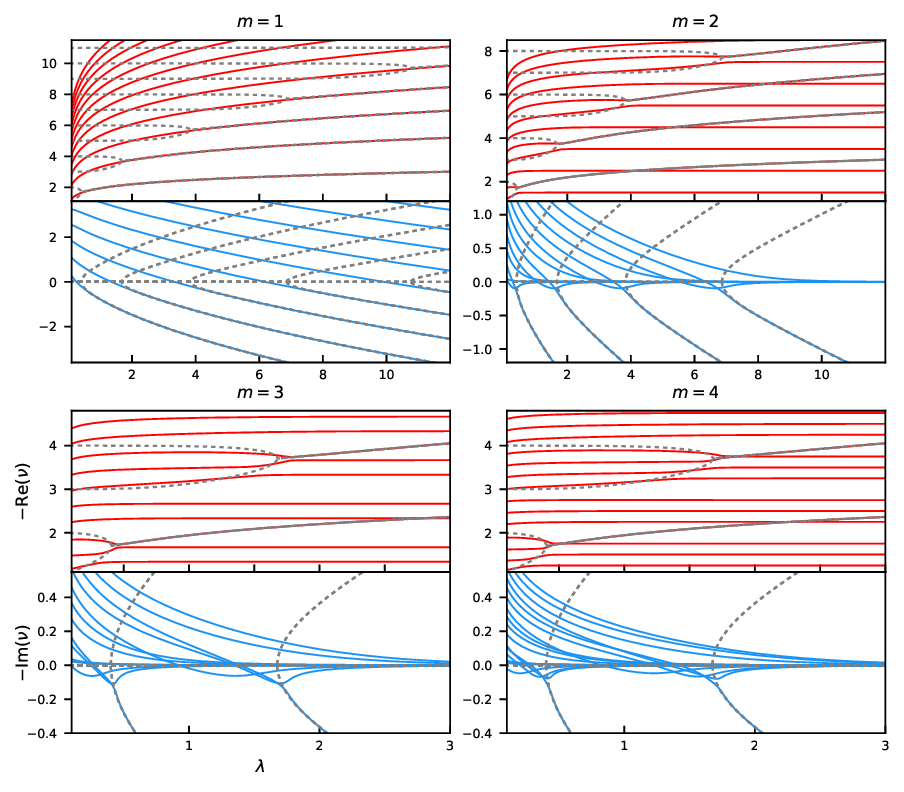}
            \caption{\label{fig:spectrum} Real and imaginary parts of the well resonances  for $m = 1, 2, 3$, and $4$ [Eq. \eqref{eq:resonance}, solid lines], and of the barrier resonances [Eq. \eqref{eq:true_resonance}, dashed lines]. }
        \end{center}
\end{figure}
It can be seen that for increasing $\lambda$, when $m = 1$, all of the well resonances converge towards the growing-state barrier ones.
For $m = 2$, half of the well resonances converge towards barrier resonances, whereas the other half appear to converge to semi-integer numbers.
Following this trend, for $m = 3$, one every three well resonances converge towards barrier resonances, wheras the other two converge towards fractions of 3 (excluding integers), and for $m = 4$, one in four well resonances converge towards barrier resonances.
It is apparent that the well resonances can be divided into two kinds; the first kind involves those solutions that converge towards barrier resonances, whereas the second kind involves those well resonances that converge towards fractional numbers, and are only present for $m > 1$.
We will not provide rigorous proof, but the existence of both kinds of solutions can somewhat be explained by analyzing Eq. \eqref{eq:resonance2}.
First, we note that,  provided that $\lambda$ is large enough,  because of Eq. \eqref{eq:asymp_K}, $K_\nu(2\sqrt{\lambda})$ is a complex number of very small magnitude. 
In addition, for ${\rm Im}(\nu_n^{(m)}) < 0$ and $m>0$, $e^{-m\nu\pi i}$ has also a very small magnitude. 
Therefore, it can be expected that the zeros of Eq. \eqref{eq:resonance} will be close to those of $\sin(m\nu\pi) \csc(\nu\pi) I_\nu(2\sqrt{\lambda})$.
The latter can occur either if $I_\nu(2\sqrt{\lambda}) = 0$, which implies that $\nu_n^{(m)} \approx \mu_n$ (this explains the first kind of resonances),  or if $\sin(m\nu\pi) \csc(\nu\pi) = 0$, which occurs whenever $\nu = k/m$, where $k$ is an integer that is not a multiple of $m$ (explaining the second kind).

%For the first kind, we note that if ${\rm Im}(\nu) < 0$ and $m > 0$, then $e^{-im\nu\pi}$ is expected to be a complex number with small absolute value.
%Therefore, the first term in the LHS of Eq. \eqref{eq:resonance} is asymptotically small when compared to ek
% since $K_\nu$ is asymptotically small [Eq. \eqref{eq:asymp_K}], the zeros of Eq. \eqref{eq:resonance} can be expected to be close to those of Eq. \eqref{eq:true_resonance}.
%Therefore, for increasing $\lambda$, the well resonances may approach the barrier ones.
%Also, since $e^{-im\nu\pi}$ becomes smaller in absolute value for increasing $m$, the convergence of the well resonances towards the barrier ones should be faster the larger $m$ is.
In Table \ref{tab:comparison}, we show $\nu_{0^*}^{(m)}$, where the index $n^*$ refers to the solution of Eq. \eqref{eq:resonance} that is closest to $\mu_n$, for several values of $\lambda$ and for increasing $m$.
\begin{table}[t]
        \begin{center}
            \begin{footnotesize}

\begin{tabular}{c|rr|r|rr}
                         \multicolumn{3}{c|}{$\lambda = 1/2$}                          &                                  \multicolumn{3}{c}{$\lambda = 2$}\\                                  
\hline
$m$      & \multicolumn{1}{c}{${\rm Re}(\nu)$} & \multicolumn{1}{c|}{${\rm Im}(\nu)$}  & \multicolumn{1}{c|}{$m$} & \multicolumn{1}{c}{${\rm Re}(\nu)$} & \multicolumn{1}{c}{${\rm Im}(\nu)$}\\
\hline
1        & -1.708889402333520                  & 0.313848239102419                     & 1                        & -2.19998150521571                   & 1.47380332298928\\                   
2        & -1.746877069032750                  & 0.289471834435959                     & 2                        & -2.19996056123564                   & 1.47382974620043\\                   
3        & -1.744703781423560                  & 0.280605627380309                     & 3                        & -2.19996055822964                   & 1.47382974508387\\                   
4        & -1.743009480401320                  & 0.281164740911281                     & 4                        & -2.19996055822965                   & 1.47382974508356\\                   
\cline{4-6}
5        & -1.743125645499290                  & 0.281441559500668                     & $\infty$                 & -2.19996055822965                   & 1.47382974508356\\                   
6        & -1.743171729577290                  & 0.281420065685151                     & \multicolumn{1}{r}{}     &                                     & \\                                   
7        & -1.743167735582120                  & 0.281412353572321                     & \multicolumn{1}{l}{}     & \multicolumn{1}{l}{}                & \multicolumn{1}{l}{}\\               
8        & -1.743166449785450                  & 0.281413091451232                     &                                  \multicolumn{3}{c}{$\lambda = 10$}\\                                 
\cline{4-6}
9        & -1.743166585008590                  & 0.281413305242469                     & \multicolumn{1}{c|}{$m$} & \multicolumn{1}{c}{${\rm Re}(\nu)$} & \multicolumn{1}{c}{${\rm Im}(\nu)$}\\
\cline{4-6}
10       & -1.743166620466710                  & 0.281413280623280                     & 1                        & -2.91772003768616                   & 4.57704829026394\\                   
11       & -1.743166616009740                  & 0.281413274758419                     & 2                        & -2.91772003768629                   & 4.57704829026402\\                   
\cline{4-6}
12       & -1.743166615042500                  & 0.281413275561234                     & $\infty$                 & -2.91772003768629                   & 4.57704829026402\\                   
13       & -1.743166615186450                  & 0.281413275720256                     & \multicolumn{1}{r}{}     &                                     & \\                                   
14       & -1.743166615212510                  & 0.281413275694549                     & \multicolumn{1}{l}{}     & \multicolumn{1}{l}{}                & \multicolumn{1}{l}{}\\               
15       & -1.743166615207940                  & 0.281413275690295                     &                                 \multicolumn{3}{c}{$\lambda = 100$}\\                                 
\cline{4-6}
16       & -1.743166615207250                  & 0.281413275691106                     & \multicolumn{1}{c|}{$m$} & \multicolumn{1}{c}{${\rm Re}(\nu)$} & \multicolumn{1}{c}{${\rm Im}(\nu)$}\\
\cline{4-6}
17       & -1.743166615207390                  & 0.281413275691218                     & 1                        & -4.32572713644441                   & 17.46003535736490\\                  
\cline{4-6}
18       & -1.743166615207410                  & 0.281413275691193                     & $\infty$                 & -4.32572713644441                   & 17.46003535736490\\                  
19       & -1.743166615207400                  & \multicolumn{1}{r}{0.281413275691190} & \multicolumn{1}{r}{}     &                                     & \\                                   
20       & -1.743166615207400                  & \multicolumn{1}{r}{0.281413275691191} & \multicolumn{1}{r}{}     &                                     & \\                                   
\cline{1-3}
$\infty$ & -1.743166615207400                  & \multicolumn{1}{r}{0.281413275691191} & \multicolumn{1}{r}{}     &                                     & \\                                   

\end{tabular}
            \end{footnotesize}
            \caption{\label{tab:comparison} $\nu_{1^*}^{(m)}$, and $\mu_0$ (denoted with $m = \infty$) for different values of $\lambda$. See Table \ref{tab:notation} for the definition of each symbol.}
        \end{center}
\end{table}
It can be seen that the distance between $\nu_{n^*}^{(m)}$ and $\mu_n$ becomes smaller with increasing $\lambda$, but also with $m$.
%The latter is due to the term $e^{-m\nu\pi i}$ also growing smaller in absolute value with $m$, which implies that the zeros of Eq. \eqref{eq:resonance} are even closer to those of $I_\nu(2\sqrt{\lambda})$.
We further demonstrate the latter by showing in Fig. \ref{fig:comparison} the computation of $-\log_{10}|\mu_n - \nu_{n^*}^{(m)}|$, for $n = 1,\ldots, 4$, and several values of $\lambda$ ranging from $1/2$ to $10$.  
It can be seen there that for fixed $\lambda$ and increasing $m$, the distance between the well and barrier resonances decreases exponentially.
In fact, one can show that
\begin{equation}
    \nu^{(m)} - \mu \approx i \, \frac{K_\mu}{\pi\dot{I}_\mu} \,e^{-i\mu\pi(2m-1)},
    \label{eq:difference}
\end{equation}
where $\dot{I}_\mu = (\partial I_\nu / \partial \nu)|_{\mu=\nu}$, by expanding the LHS of Eq. \eqref{eq:resonance} a Taylor series about $\nu = \mu$ [where $\mu$ is a root of Eq. \eqref{eq:true_resonance}], keeping the linear term, and assuming that the term involving $\dot{I}_\mu$ is much larger (in the asymptotic sense) than those involving $K_\mu$ and $\dot{K}_\mu$.
Although Eq. \eqref{eq:difference} is not proved rigorously, we have performed several computations that confirm it numerically.

Going back to the barrier problem [Eq. \eqref{eq:schrodinger}], one may wonder if a similar set of solutions such as those of Eq. \eqref{eq:resonance} may be discovered through complex rotation, since if $r = \rho e^{i\theta}$ is substituted in the definition of $t$, it follows that ${\rm Arg}\,t = -\rho \sin (\theta)/2$; i.e., ${\rm Arg}\,t$ changes with $\rho$, thereby modifying the asymptotic behavior of the solution depending on the region where the boundary condition is imposed, just as in the case of the well problem.
The answer is that it would not, since, even though an equation analogous to \eqref{eq:convergence_region} might be posed for the barrier problem, the analytical continuation of $I_\nu$ is proportional to the same function [see \eqref{eq:cont_I}]; the set roots of an equation equivalent to \eqref{eq:resonance} are then the same for every $m$.

\section{Riccati-Pad\'e method} \label{sec:RPM}

The purpose of the present section is to apply the RPM to both Eqs. \eqref{eq:schrodinger} and \eqref{eq:schrodinger2}. 
The application of the RPM to central-force problems was presented in Ref. \onlinecite{Fernandez_1989}, but for completeness we briefly review it here.

The RPM can be applied to any central-force problem, i.e.,
\begin{equation}
    -\psi^{\prime\prime}(r) + \left[V(r) + \frac{l(l+1)}{r^2}- E\right] \psi(r) = 0,
    \label{eq:central_force}
\end{equation}
with $\psi(0) = 0$, and $\psi(\infty) = 0$, where the potential $V(r)$ can be expanded in a Laurent series of the form $\sum_{j=-1}^{\infty} v_j r^j$. 
In that case, $\psi(r) \sim C r^{l+1}$ for $r \rightarrow 0$, and its logarithmic derivative is singular at origin, $\psi^\prime(r) / \psi(r) \sim (l+1)/r$; therefore it must be regularized. 
The resulting regularized function, $f(r) = (l+1)/r - \psi^{\prime}(r) / \psi(r)$, can then be expanded in a Taylor series about the origin, $f(r) = \sum_{j=0}^{\infty} f_j r^j$.
It is a solution to the Riccati equation,
\begin{equation}
    f^\prime(r) -f^2(r) + \frac{2(l+1)}{r}f(r) - \left[E - V(r)\right] = 0,
    \label{eq:riccati}
\end{equation}
from which a recurrence relation can be found for the $f_j$ coefficients,
\begin{equation}
    f_{j+1} = \frac{1}{2l+j+3} \left[\sum_{i=0}^{j}f_i f_{j-i} - v_j + E \delta_{j0}\right],
    \label{eq:fcoef}
\end{equation}
with $f_0 = -v_{-1}/(2l+2)$.
$f(r)$ can also be approximated by a Pad\'e approximant of order $[M/N]$, i.e., a quotient of polynomials of orders $M$ and $N$ that are typically chosen in such a way that $[M/N](r) - f(r) = \mathcal{O}(r^{M+N+1})$.
If the condition $[M/N](r) - f(r) = \mathcal{O}(r^{M+N+2})$ is imposed instead, it leads to a system of equations that has nontrivial solutions if
\begin{equation}
    |H(D,d)| = \begin{vmatrix}
        f_{d+1} & f_{d+2} & \ldots & f_{d+D} \\
        f_{d+2} & f_{d+3} & \ldots & f_{d+D+1} \\
        \vdots &  \vdots  & \ddots & \vdots \\
        f_{d+D} & f_{d+D+1} & \ldots & f_{2D+d-1}
    \end{vmatrix} = 0.
    \label{eq:hankdet}
\end{equation}
Here we have defined $d = M-N$ and $D = N+1$.
The Hankel determinants [the determinants of $H(D,d)$] can be computed efficiently by resorting to the following recurrence relation, 
\begin{equation}
    H_D^d = \frac{H_{D-1}^{d} H_{D-1}^{d-2} - (H_{D-1}^{d+1})^2 }{H_{D-2}^{d-2}},
    \label{eq:hankdet_recursion}
\end{equation}
 where $H_D^d = |H(D,d)|$, with $H_0^d = 1$.

The RPM is closely related to the process of sending a movable pole at infinity \cite{Abbasbandy_2011}.
This pole can be moved through any ray in the complex plane, and for this reason the solutions of Eq. \eqref{eq:hankdet} yield approximations to the eigenvalues of Eq. \eqref{eq:central_force} whose corresponding eigenfunctions that asymptotically decay through different regions of the complex plane.
Concordantly, the roots of Eq. \eqref{eq:hankdet} persist if a complex rotation of the variable is performed \cite{Fernandez_2016_a}.
Another important characteristic of the RPM is that $H_D^d$ typically exhibits a great number of roots in the neighborhood of each eigenvalue; this makes it easier to differentiate the roots that approximate an eigenvalue from those that are spurious, but it also complicates the task to find the optimal sequence of roots that converges to each eigenvalue by means of iterative methods.

\subsection*{Application to the present problem}

In the present work we have employed two different approaches to compute the roots of Eq. \eqref{eq:hankdet}.
The first of them is to compute an expression for $H_D^d$ analytically, and solve for $E$ using a polynomial root-finding algorithm; in this way, all of the roots of $H_D^d$ can be found for a given $D$, but it is only applicable to small $D$ values, since the degree of $H_D^d$ grows rapidly.
The second one is to use the recurrence relation \eqref{eq:hankdet_recursion} to compute $H_D^d$ and its derivative numerically, and to employ the Newton-Raphson (NR) to find its roots.
It has the advantage of being extremely efficient, but only one root can be computed at a time, and a good initial value of the energy is required for the NR method to converge.
Since the Hankel determinants present clusters of roots that approach the eigenvalues, it is not guaranteed that the NR method will find the optimal one, but with large $D$, even the suboptimal sequences of roots provide a good approximation to the eigenvalues.
We are typically able to compute determinants with $D \leq 30$ with the first approach before running out of memory, whereas with the second one, we have reported results with $D > 800$ \cite{Fernandez_2017}, and even larger determinants are possible, provided that the starting points for the NR method are sufficiently accurate.
We would like to mention here that we have utilized the GiNaC library \cite{Bauer_2002, Vollinga_2006} for the implementation of the second approach.

By setting $V(r) = \lambda e^{\pm r}$, and $l = 0$, Eq. \eqref{eq:central_force} yields Eq. \eqref{eq:schrodinger} or \eqref{eq:schrodinger2}.
The choice of the sign in $V(r)$ is irrelevant since, as stated earlier, the roots of $H_D^d$ remain unmodified.
Since the results obtained by the RPM do not typically change with $d$, we have set $d = 0$ in all of our computations.
We have computed all of the roots of $H_{30}^0$, and selected the ones that coincide with the exact eigenvalues of both Eqs. \eqref{eq:schrodinger} and \eqref{eq:schrodinger2}; these are shown, for $\lambda = 1/2$, in Table \ref{tab:RPM1}, and, for $\lambda = 10$, in Table \ref{tab:RPM2}.
It can be appreciated that some of them approach the bound states, and others approach the barrier and well resonances, but none of them are close to any of the virtual states.
For both values of $\lambda$, the bound states and resonances corresponding to $m = 1$ are found with more significant digits than those with $m = 2$, and in the latter case, there is also one root that corresponds to an $m = 3$ resonance.
Among the results presented in Table \ref{tab:RPM1} is included the seemingly wrong ground state for the exponential barrier, $E \approx -0.705450568055028 + 0.268165964871580\,i$, found in Ref. \onlinecite{Amore_2008}.
This value coincides with $-[\nu_{1^*}^{(1)}]^2/4$, where $\nu_{1^*}^{(1)}$ is taken from Table \ref{tab:comparison}.
The first root with $m=1$, and the second one with $m=2$ of Table \ref{tab:RPM1} may also be considered as approximations towards $\varepsilon_0(\lambda=1/2) \approx -0.73985-0.2452i$, whereas the third entries of Table \ref{tab:RPM2} for $m = 1, 2$ may be considered as approximations towards $\varepsilon_0(\lambda=10) \approx 3.10907-6.6772i$.
We also note here that the roots of Eq. \eqref{eq:hankdet} are either real or come in complex conjugate pairs, since the $v_j$ coefficients of Eq. \eqref{eq:fcoef} are real, the $f_j$ coefficents are polynomials on $E$ with real coefficients, and so is $H_D^d$; therefore, the same analysis performed here is valid for the growing-state solutions.

\begin{table}[t]
    \begin{footnotesize}
        \begin{center}
            \begin{tabular}{lD{.}{.}{20}D{.}{.}{20}}
                \multicolumn{1}{c}{$m$} & \multicolumn{1}{c}{${\rm Re}(\epsilon)$} & \multicolumn{1}{c}{${\rm Im}(\epsilon)$} \\
                \hline
                0        & 3.2292159472536048534          & 0.0                            \\
                0        & 7.0837240758343799367          & 0.0                            \\
                0        & 11.541280962123897449          & 0.0                            \\
                0        & 16.53405783500778465           & 0.0                            \\
                0        & 27.94592416939                 & 0.0                            \\
                0        & 41.0616                        & 0.0                            \\
                0        & 48.206                         & 0.0                            \\
                1        & -0.70545056805502837410        & 0.26816596487157970576         \\
                1        & -2.0145028385826182272         & -0.91594563985411650792        \\
                1        & -3.17882146596656262           & -2.84627255601162901           \\
                1        & -4.19709693935568              & -5.32673986864807              \\
                1        & -5.07091771510                 & -8.26819182266                 \\
                1        & -5.80277763                    & -11.6168375                    \\
                1        & -6.395162                      & -15.33562                      \\
                1        & -6.850                         & -19.40                         \\
                2        & -0.0624600582                  & -0.00235480490                 \\
                2        & -0.7419463                     & 0.2528359                      \\
                2        & -0.5549276                     & 0.04161432                     \\
                2        & -1.522                         & -0.1790                        \\
                \hline
            \end{tabular}
        \end{center}
    \end{footnotesize}
    \caption{\label{tab:RPM1} Complex eigenvalues obtained with $H_{30}^0 = 0$, for $\lambda = 1/2$. Results coincide with the exact ones up to the last digit reported. They have been truncated at 20 significant digits.}
\end{table}
\begin{table}[t]
    \begin{footnotesize}
        \begin{center}
            \begin{tabular}{lD{.}{.}{20}D{.}{.}{20}}
                \multicolumn{1}{c}{$m$} & \multicolumn{1}{c}{${\rm Re}(\epsilon)$} & \multicolumn{1}{c}{${\rm Im}(\epsilon)$} \\
                \hline
                0        & 24.095880341888706123          & 0.0                            \\
                0        & 39.078924487590608756          & 0.0                            \\
                0        & 54.342383405484864190          & 0.0                            \\
                0        & 70.173203793408790845          & 0.0                            \\
                0        & 86.630901797549645582          & 0.0                            \\
                0        & 103.722769025200               & 0.0                            \\
                0        & 121.4400060068                 & 0.0                            \\
                0        & 139.768519                     & 0.0                            \\
                0        & 158.693                        & 0.0                            \\
                1        & -10.243001240262833149         & 6.9319297923706305098          \\
                1        & -3.7481264282889514022         & 8.0918634524566962810          \\
                1        & 3.1090702082731894910          & 6.6772727549801459614          \\
                1        & -16.525883780702648862         & 4.1357639513958152182          \\
                1        & -22.63511276702947437          & 0.09542171991804396736        \\
                1        & -28.583717704014               & -4.97259427564364              \\
                1        & -34.3769116736                 & -10.93053096967                \\
                1        & -40.017132                     & -17.682645                     \\
                1        & -45.506                        & -25.1582                       \\
                2        & -0.0624998                     & -0.000000526971                \\
                2        & -3.74812643                    & 8.09186345                     \\
                2        & 3.109070208273                 & 6.677272754981                 \\
                3        & -3.748126                      & 8.091863                       \\
                \hline
            \end{tabular}
        \end{center}
    \end{footnotesize}
    \caption{\label{tab:RPM2} Complex eigenvalues obtained with $H_{30}^0 = 0$, for $\lambda = 10$. Results coincide with the exact ones up to the last digit reported. They have been truncated at 20 significant digits.}
\end{table}

\begin{figure}[t]
        \begin{center}
            \includegraphics[]{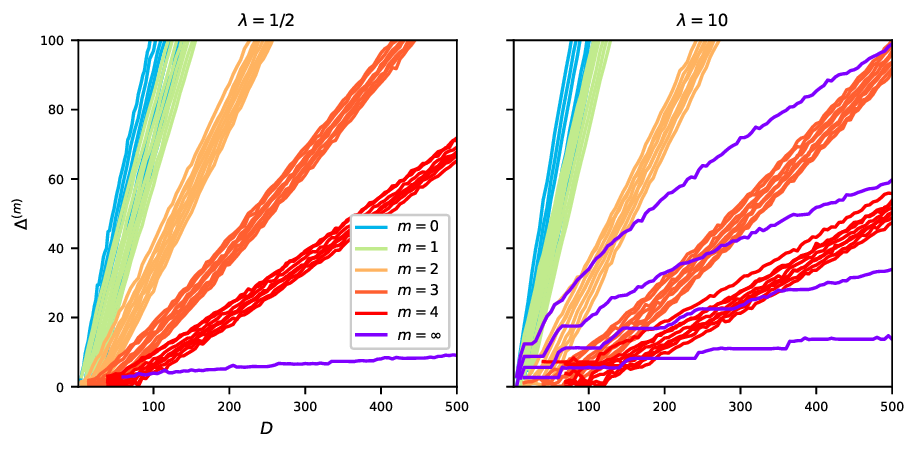}
            \caption{\label{fig:RPM} $\Delta_n^{(m)}(D)$ for $\lambda = 1/2$ and $\lambda= 10$. $m = \infty$ denotes $\Delta_n(D)$, instead. }
        \end{center}
\end{figure}

It is expected that for increasing $D$, the accuracy of these results will improve, but also that there will be roots of $H_D^0$ that approach the well resonances with $m \geq 4$.
Since the computation of the whole set of roots of $H_D^d$ becomes expensive for $D > 30$, in order to study how rapidly the roots of $H_D^d$ converge towards both the barrier and well resonances, we have resorted to our NR-based approach. 
With it, we have computed $\tilde{\epsilon}_n^{(m)}$, which is the root of $H_D^0$ closest to $\epsilon_n^{(m)}$, and $\tilde{\varepsilon}_n$, the root of $H_D^0$ closest to $\varepsilon_n$, for $D = 5, 10, \ldots, 500$, $m = 0,\ldots,4$, $n = 0,\ldots, 7$, and $\lambda = 1/2, 10$.
We have employed the exact resonance energies computed from Eqs. \eqref{eq:true_resonance} and \eqref{eq:resonance} as a starting point for the NR method. 
Instead of showing the tens, or even hundreds, of coincident significant digits between the exact eigenvalues and their RPM approximations, we illustrate how accurate the latter are by means of the quantities $\Delta_n^{(m)}(D) = -\log_{10} | \tilde{\epsilon}_n^{(m)}(D) - \epsilon_n^{(m)}|$, and $\Delta_n(D) = -\log_{10} | \tilde{\varepsilon}_n(D) - \varepsilon_n|$, both of which provide a good estimate of the number of correct decimal places. 
The plots of $\Delta_n^{(m)}(D)$ and $\Delta_n(D)$ are shown in Fig. \ref{fig:RPM}; the latter labeled as $m = \infty$.
It can be seen there that $\Delta_n^{(m)}(D)$ are straight lines whose slopes depend on $m$, but not on $n$.
The position of these straight lines does depend on $n$, and we have found that in general they move to the right when $n$ increases.
Both findings are not surprising: on one hand, for increasing $n$, the exact eigenfunction is expected to have a larger number of nodes, which increments the number of $f_j$ coefficients that are needed to represent its logarithmic derivative correctly. 
On the other, for increasing $m$, the boundary conditions are moved further to the right (i.e., to larger $|r|$), which also increases the number of coefficients needed.
As is made apparent by our computations, the effect of the latter is more noticeable.
The plots of $\Delta_n(D)$ show that the roots of $H_D^d$ also converge towards the barrier resonances, albeit more slowly.
Instead of being straight lines, they are curves that are concave everywhere, except for a number of plateaus which is different for each state.
Convergence rate is lower for higher $n$, meaning that the top curve in the $\lambda = 10$ plot corresponds to $\Delta_0$, the curve below it corresponds to $\Delta_1$, and so on.
To ease the discussion of these results, we have compiled some logarithmic differences between well and barrier resonances, i.e., $-\log_{10}|\nu_{n^*}^{(m)} - \mu_n|$, in Table \ref{tab:deltas}.

Focusing first on the curve for $\Delta_0$, we note that, when $D < 15$, it coincides with $\Delta_{0^*}^{(1)}$, which suggests that the solution being obtained is a well resonance, whose eigenvalue coincides with that of a barrier resonance up to approximately 12 decimal places.
At $D = 15$, a small region is reached where incrementing $D$ does not seem to improve accuracy; in that region, $\Delta_0 \approx 12.5$ coincides with $-{\rm log}_{10} | \varepsilon_0 - \epsilon_{0^*}^{(1)}| $, implying that $\tilde{\epsilon}_{0^*}^{(1)}$ is a more accurate representation of $\epsilon_{0^*}^{(1)}$ than $\epsilon_{0^*}^{(1)}$ is of $\varepsilon_0$.
When $D > 25$, convergence seems to pick up, and the plot of $\Delta_0$ becomes a concave curve with no additional plateaus; i.e., for $D > 25$ and $m > 1$, $\Delta_0 > \Delta_{0^*}^{(m)}$.
The curve for $\Delta_1$ is similar, but presenting two plateaus; before the first plateau, it coincides with the curve for $\Delta_{1^*}^{(1)}$. 
Then, at the first plateau, located approximately between $D = 15$ and $D = 40$, $\Delta_{1} = 8.75$, in coincidence with the entry for $m = 1$, $n = 0$ of Table \ref{tab:deltas}.
After that, $\Delta_1$ adopts the form of $\Delta_{1^*}^{(2)}$, until it reaches a second plateau where $\Delta_1 \approx 17.53$ [c.f. Table \ref{tab:deltas}], located in a region where $65 < D < 80$.
Finally, it assumes its concave shape, where $\Delta_1 > \Delta_{1^*}^{(m)}$, for $m > 2$.
Similar explanations can be devised for the curves for $\Delta_2$, which seems to present four plateaus (all included in Table \ref{tab:deltas}), and for $\Delta_3$, which shows about 5 plateaus, all of them also included in the same table.
The plateaus observed in Fig. \ref{fig:RPM} are quite probably an artificial byproduct of our method to find the roots of $H_D^d$.
It is very likely that there are sequences of roots of the Hankel determinants that coincide with the concave part of the curves we described, but do not have the plateaus nor the parts that match exactly with the curves for well resonances; we are just not able to find them since we do not have good starting points for the NR method.

\begin{table}[b]
    \begin{center}
        \begin{tabular}{c|rrrrr}
            \multirow{2}{*}{$m$} & \multicolumn{4}{c}{$n$} \\
            \cline{2-5}
                                 & \multicolumn{1}{c}{0} & \multicolumn{1}{c}{1} & \multicolumn{1}{c}{2} & \multicolumn{1}{c}{3}\\
    \hline
    1               & 12.39           & 8.75            & 5.60            & 2.70           \\
    2               & 24.88           & 17.53           & 11.22           & 5.46           \\
    3               & 37.37           & 26.30           & 16.85           & 8.22           \\
    4               & 49.86           & 35.07           & 22.47           & 10.97          \\
    5               & 62.34           & 43.85           & 28.10           & 13.73          \\
        \end{tabular}
    \end{center}
    \caption{\label{tab:deltas} $-\log_{10}| \varepsilon_n - \epsilon_{n^*}^{(m)}|$ for the four states with $\lambda = 10$.}
\end{table}

The results presented above suggest that the roots of the Hankel determinants converge towards both the well and barrier resonances.
The authors of Ref. \onlinecite{Amore_2008} only found the well resonances, since they performed computations with $D \leq 20$, which is not enough to discover the barrier ones.

\section{Conclusions} \label{sec:conclusions}

We have studied the analytical solutions to the Schr\"odinger equation with a repulsive exponential potential, as well as those to the Schr\"odinger equation with an infinite exponential well, and shown that the complex spectra associated to both problems are related as some of the eigenvalues of the latter tend to those of the former when the potential parameter $\lambda$ increases.
Those well resonances that do not tend towards barrier ones do so instead to real fractional numbers.
Starting from any of the two problems, if a complex rotation $r = \rho e^{i\theta}$ is performed, then $\theta_{\rm crit} = \pi/2$ separates one set of resonances from the other.
These findings are similar to those of Refs. \onlinecite{Rittby_1981, Rittby_1982, Korsch_1982, Rittby_1982_a} regarding the potential proposed by Moiseiev\cite{Moiseyev_1978}, which had two sets of resonances that were uncovered by complex rotation, separated by $\theta_{\rm crit} = \pi/4$.
In both cases, even though the asymptotic behavior of the solutions is markedly different, the eigenvalues remain quite close to each other.

Regarding the RPM, we have found a satisfactory explanation for the seemingly wrong results obtained in Refs. \onlinecite{Fern_ndez_1995, Fern_ndez_1996, Amore_2008}: all the resonances that are deemed wrong there are in fact well resonances.
We have computed the roots of the Hankel determinants and analyzed how they converge towards both the well and barrier resonances, our findings suggesting that there are sequences of them that converge independently to both of their kinds.
Contrarily, we were not able to obtain any of the virtual states by means of this method.
All of the above suggests that it is only possible to obtain approximations to eigenvalues that are discoverable by a complex rotation, even though the RPM does not require to perform one explicitly.
The results obtained in the present work therefore provide evidence that the convergent sequences of roots of the Hankel determinants always correspond to eigenfunctions that decay asymptotically through some path in the complex plane.

\section*{Acknowledgments}

This work was funded by Consejo Nacional de Investigaciones Cient\'ificas y T\'ecnicas (CONICET).

\section*{Conflict of interest statement}

The author has no conflicts to disclose.

\section*{Data availability}

The data that support the findings of this study are available from the corresponding author upon reasonable request.

%\begin{acknowledgments}
% Put your acknowledgments here.
%\end{acknowledgments}

\bibliography{expo.bib}

\end{document}